\documentclass[%
 reprint,
 amsmath,amssymb,
 aps,
]{revtex4-2}

\usepackage{color}
\usepackage{tabularx}
\usepackage{graphicx}
\usepackage{dcolumn}
\usepackage{bm}
\usepackage{amsmath, amssymb, amsthm}
\usepackage[normalem]{ulem}
\usepackage{graphicx}
\usepackage{braket}
\usepackage{booktabs}
\usepackage{multirow}
\usepackage{tikz}
\usetikzlibrary{positioning, shapes.geometric, arrows}
\usepackage{qcircuit}

\begin{document}

\preprint{APS/123-QED}

\title{\Large{Scalable On-Hardware Training of Quantum Neural Networks \\ and Application to Clinical Data Imputation}}

\author{Natansh Mathur}
\email{nashmathur@gmail.com}
\affiliation{IRIF, CNRS and Université Paris Cité}
\affiliation{QC Ware, France}

\author{Panagiotis Kl. Barkoutsos}%
\affiliation{IonQ}

\author{Masako Yamada}%
\affiliation{IonQ}

\author{Iordanis Kerenidis}%
\affiliation{IRIF, CNRS and Université Paris Cité}
\affiliation{Quantum Signals}

\author{Martin Roetteler}%
\affiliation{IonQ}


\begin{abstract}

Training quantum neural networks (QNNs) on quantum hardware is currently bottlenecked by the cost of gradient estimation: standard parameter-shift methods require a number of circuit evaluations that grows quadratically with the number of trainable parameters, making hardware-based optimisation impractical beyond small system sizes. In this work, we introduce a training framework that reduces this cost to logarithmic in the number of qubits, making gradient-based QNN optimisation feasible on near-term hardware at increasing scales.

Our framework combines three co-designed ingredients: (i) a structured, subspace-preserving Butterfly circuit architecture with $O(n \log n)$ parameters and logarithmic depth; (ii) a layer-wise training strategy that confines on-hardware optimisation to one small, well-structured layer at a time; and (iii) a parallelised parameter-shift rule that exploits the commuting structure within each Butterfly layer to extract all gradients in a constant number of circuit executions. Together these reduce the number of distinct circuit evaluations per optimisation step from $O(n^2)$ to $O(\log n)$.

We validate the framework on clinical data imputation using the MIMIC-III electronic health record dataset, a demanding benchmark sensitive to optimisation instability and model variance. Hybrid classical–quantum models are trained directly on IonQ Forte Enterprise trapped-ion hardware at 16 qubits without performance degradation relative to ideal or noisy simulation and via tensor-network simulation at 32 qubits, with 32-qubit inference executed on hardware. The resulting models match or exceed strong classical neural baselines in downstream patient survival prediction while exhibiting reduced variance across runs, demonstrating that the proposed framework enables practical, scalable QNN training under realistic hardware constraints.

\end{abstract}

\maketitle


\section{\label{sec:intro}Introduction}



Quantum machine learning (QML) aims to leverage quantum processors for learning,
optimisation, and data-driven modelling tasks that are challenging for classical methods.
A central model class in this programme is the quantum neural network (QNN): a
parametrised quantum circuit trained via classical feedback that defines a hypothesis
space in a high-dimensional Hilbert space~\cite{schuld2019evaluating}.
Despite substantial theoretical and experimental progress, the practical utility of QNNs
remains constrained by fundamental challenges of trainability and hardware
execution challenges.
These bottlenecks are sharply exposed by \emph{data imputation}, where missing
values must be inferred from partially observed data.
In clinical and biomedical datasets, missingness arises routinely due to heterogeneous
measurement schedules, acquisition failures, and data-entry issues, and robust imputation
is essential for reliable downstream prediction and decision support~\cite{rubin2018multiple,
sterne2009multiple}.
Imputation tasks are sensitive to optimisation instability, model variance, and noise, especially when feature dependencies are high-dimensional and non-linear,
making clinical imputation a stringent and informative testbed for QML models operating
under realistic resource constraints.

From the perspective of QML, data imputation poses two intertwined challenges. 
First, effective imputers must learn conditional relationships in moderately high-dimensional feature spaces, placing demands on model expressivity. 
Second, they must do so reliably under limited data, noise, and stochastic optimisation, making training stability as important as peak predictive accuracy. 
These requirements align closely with the central obstacles currently facing QNNs, and therefore provide a natural application domain in which to assess whether training-efficient quantum models can function as reliable components of real-world data analysis pipelines.

Two bottlenecks constrain QNN trainability on near-term and early fault-tolerant quantum hardware. 
The first is the prevalence of barren plateaus, where gradients of common cost functions vanish exponentially with system size or circuit depth, rendering gradient-based optimisation ineffective~\cite{mcclean2018barren, cerezo2021cost}; Hamming-weight-preserving circuits mitigate this by confining evolution to fixed-excitation subspaces with provably favourable gradient scaling~\cite{monbroussou2025trainability}.
The second, and more immediate for hardware execution, is the cost of gradient estimation: the parameter-shift rule yields exact, hardware-compatible derivatives~\cite{mitarai2018quantum, schuld2019evaluating}, but for generic $n$-qubit architectures with $O(n^2)$ trainable parameters, each optimisation step requires $O(n^2)$
distinct circuit evaluations which is a requirement that is the dominant experimental bottleneck on NISQ devices~\cite{abbas2023quantum}. 
The first experimental on-hardware QNN training via parameter shift, QOC~\cite{Wang2022}, inherited this scaling and required probabilistic gradient pruning to recover accuracy from hardware noise; more recent proposals such as weighted stochastic block descent (WSBD)~\cite{kverne2026wsbd} dynamically freeze low-importance parameters to reduce circuit evaluations per step, but retain the fundamental per-parameter shift cost.
The training framework introduced in this work breaks from this trajectory: by co-designing the Butterfly circuit architecture, a layer-wise training protocol, and a parallelised parameter-shift rule that exploits commuting structure within each layer, we reduce the number of distinct circuit evaluations per optimisation step from $O(n^2)$ to $O(\log n)$, enabling direct gradient-based training on IonQ Forte Enterprise at 16 qubits and 32-qubit inference on hardware.

A route towards improved trainability is to impose structure on the circuit architecture by restricting its dynamics. Hamming-weight-preserving QNNs constrain evolution to fixed-excitation subspaces of the Hilbert space, reducing the effective dimension explored during optimisation while retaining meaningful modelling capacity~\cite{landman2022quantum, kerenidis2022quantum}; combined with an explicit data-loading layer, the choice of encoding and circuit depth provides a controllable trade-off between expressivity and trainability~\cite{havlivcek2019supervised, schuld2018supervised, thakkar2024improved, monbroussou2025trainability}. 
Within this class, the Butterfly architecture is especially attractive for hardware execution: logarithmically many layers of two-qubit, excitation-preserving gates in a fast-transform-inspired pattern~\cite{cherrat2024quantum} realise global information mixing with shallow depth, reducing trainable parameters from $O(n^2)$ to $O(n \log n)$, while gates within each layer act on disjoint qubit pairs, enabling parallel execution and aligning naturally with long-range connectivity platforms such as trapped-ion processors.



Even with these architectural advantages, naively training all parameters simultaneously on hardware remains impractical. 
We therefore combine two complementary strategies. 
First, \emph{layer-wise training} exploits the hierarchical structure of the Butterfly architecture: rather than optimising all parameters jointly, we train progressively, fixing learned layers and introducing new ones sequentially. 
With only $O(\log n)$ layers, this confines on-hardware optimisation to a small, well-structured parameter subset at each stage, reducing experimental overhead, a strategy well established in classical deep learning and block-coordinate optimisation~\cite{hinton2006fast, tseng2001convergence}. 
Second, \emph{parallelised parameter-shift} exploits the commuting structure within each Butterfly layer: gates act on disjoint qubit pairs with mutually commuting generators, so all parameters in a layer can be shifted simultaneously while individual gradients remain extractable, making the per-layer circuit cost a small constant independent of layer size~\cite{bowles2023backpropagation, coyle2024training}. 
Together, these strategies reduce the number of distinct circuit evaluations per optimisation step from $O(n^2)$ in generic architectures to $O(\log n)$, compatible with the measurement, noise, and wall-clock constraints of current quantum hardware.

In this work, we demonstrate the practical relevance of this training-efficient framework by applying it to clinical data imputation. 
We use a hybrid classical–quantum pipeline in which a QNN serves as a learnable conditional estimator within an iterative-imputation-style procedure, while the remaining components are handled classically. 
Imputation quality is assessed via downstream patient survival prediction, providing a clinically meaningful and application-relevant performance measure. 
We evaluate the approach using a combination of quantum hardware execution and simulation at system sizes of 8, 16, and 32 qubits. 
The results show that QNN components can be trained directly on trapped-ion hardware without performance degradation relative to simulation, and that hybrid classical–quantum imputers achieve performance comparable to classical neural baselines while exhibiting improved robustness across runs in a controlled one-feature imputation protocol.

\section{Training QNNs on Quantum Hardware}
\label{sec:training_qnns}

Training parametrised quantum circuits directly on near-term quantum processors remains one of the central practical obstacles to quantum machine learning. While quantum neural networks (QNNs) can in principle represent rich hypothesis classes, their optimisation on hardware is typically constrained by (i) limited coherence and gate fidelity, (ii) finite shot budgets and wall-clock time, and (iii) the high experimental overhead of gradient estimation required by classical outer-loop optimisation \cite{mitarai2018quantum, schuld2019evaluating, abbas2023quantum}. These constraints are particularly acute for architectures with many trainable parameters, where standard gradient estimators require executing a large number of distinct circuits per optimisation step.

In this section, we present a hardware-oriented training framework designed to make gradient-based QNN optimisation feasible at increasing system sizes. The framework combines three ingredients that are co-designed to match experimental constraints.
First, we employ a structured, subspace-preserving QNN architecture built from Hamming-weight preserving two-qubit gates. This restriction improves trainability by confining dynamics to fixed-excitation subspaces and by enabling controlled expressivity through the choice of data encoding and circuit depth \cite{landman2022quantum, kerenidis2022quantum, havlivcek2019supervised, schuld2018supervised, thakkar2024improved}. Concretely, our trainable ansatz uses a Butterfly layout of logarithmic depth, which realises global mixing while reducing the parameter count from the $O(n^2)$ scaling typical of dense two-qubit layers to $O(n\log n)$ for an $n$-qubit model \cite{cherrat2024quantum}.

Second, we adopt a layer-wise training strategy tailored to the hierarchical Butterfly structure. Rather than optimising all circuit parameters simultaneously, we train progressively: previously learned layers are fixed while new layers are introduced and optimised sequentially. This approach reduces the effective optimisation dimension at each stage, improves stability under hardware noise, and is well matched to Butterfly circuits because they contain only $O(\log n)$ layers. It also supports hybrid workflows in which earlier layers are trained using classical resources or quantum simulation and are then held fixed when moving optimisation to hardware for selected layers. Such progressive or greedy layer-wise optimisation schemes are well established in classical deep learning and block-coordinate optimisation \cite{hinton2006fast,tseng2001convergence}, providing additional justification for their use in the quantum setting.

Third, and crucially for experimental efficiency, we exploit the commuting structure within each Butterfly layer to enable parallel gradient evaluation. In the Butterfly ansatz, the two-qubit gates within a given layer act on disjoint qubit pairs; their generators, therefore, commute. For commuting-generator circuits, one can shift all parameters in a layer simultaneously while still extracting the individual partial derivatives, enabling a parallelised parameter-shift rule that reduces the number of circuit evaluations required for a full-layer gradient from scaling with the number of parameters to a constant number of circuit executions \cite{bowles2023backpropagation, coyle2024training}. Combined with layer-wise training, this implies that each optimisation step requires only a constant number of circuit executions per layer, and hence $O(\log n)$ distinct circuit evaluations per step for an $n$-qubit Butterfly QNN.

The remainder of this section instantiates this framework in detail. We first describe the QNN architecture, including non-Gaussian state initialisation, data loading, and the subspace-preserving Butterfly ansatz. We then formalise the layer-wise hybrid training protocol and derive the parallel parameter-shift estimator used for efficient gradient evaluation on hardware. Together, these components yield a training pipeline compatible with the measurement, noise, and wall-clock constraints of current trapped-ion processors, and they will be used in subsequent sections to support the end-to-end hybrid imputation experiments.

\subsection{Quantum Neural Network Architecture}
\label{subsec:qnn_architecture}

The quantum neural network (QNN) architecture employed in this work is designed around three guiding principles:
(i) controllable expressivity to mitigate barren plateaus,
(ii) favourable parameter and depth scaling for hardware execution, and
(iii) structural compatibility with efficient gradient evaluation.
To this end, the model is organised as a modular pipeline consisting of a non-Gaussian state initialisation stage, a data-loading layer, and a parametrised, subspace-preserving quantum circuit arranged in a Butterfly architecture. This modular structure allows each component to be independently motivated and optimised, while ensuring coherent end-to-end operation on quantum hardware.


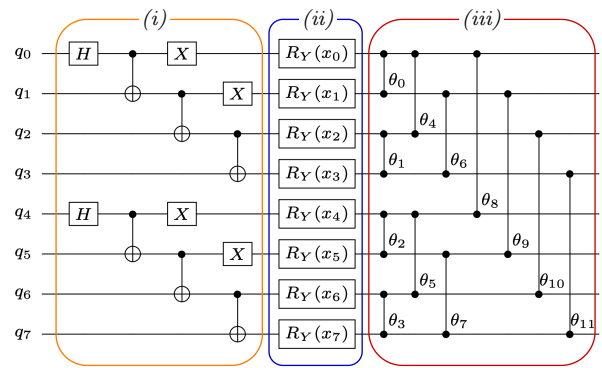
\begin{figure}[!tb]
    \centering
    \begin{tikzpicture}
    \node[anchor=south west, inner sep=0] (image) at (0,0) {%
        \includegraphics[width=\columnwidth]{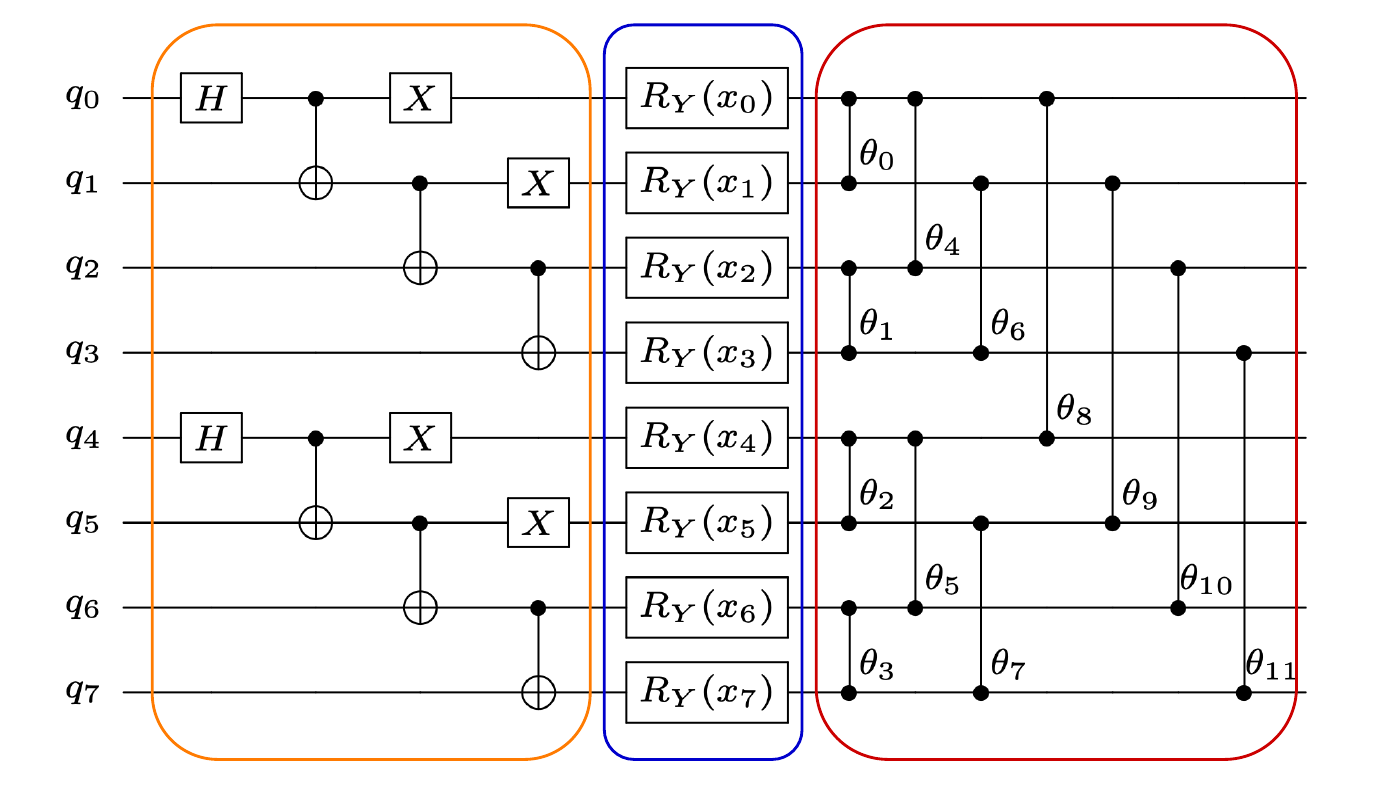}%
    };
    \begin{scope}[x={(image.south east)}, y={(image.north west)}]
        \node[font=\footnotesize\itshape, fill=white, inner sep=1pt] at (0.26, 0.965) {(i)};  
       \node[font=\footnotesize\itshape, fill=white, inner sep=1pt] at (0.51, 0.965) {(ii)};  
        \node[font=\footnotesize\itshape, fill=white, inner sep=1pt] at (0.76, 0.965) {(iii)}; 
    \end{scope}
\end{tikzpicture}

 
  \caption{The quantum component of the hybrid imputation pipeline, shown for $n = 8$
qubits, consists of three modules. \textit{(i)~Non-Gaussian state initialisation}: each group
of four qubits is prepared in the entangled state $|\phi\rangle = (|0011\rangle\!+\!|1100\rangle)/\sqrt{2}$ using the magic-state loader of~\cite{oszmaniec2022fermion}, applied in parallel across $n/4$ qubit groups; this places the register in a non-Gaussian, Hamming-weight-preserving state that is inaccessible to purely Gaussian dynamics and classically hard to simulate. 
\textit{(ii)~Data-loading layer}: each classical feature $x_i$ is angle-encoded via a single-qubit rotation $R_Y(2\pi x_i)$ applied to qubit $i$, acting on top of the non-Gaussian initial state.
\textit{(iii)~Trainable Butterfly circuit}: $O(\log n)$ layers of two-qubit RBS gates with trainable parameters $\boldsymbol{\theta}$ realise global information mixing within the
Hamming-weight-preserving subspace, keeping the total parameter count at
$O(n \log n)$.
In our experiments, the quantum model serves as a learnable conditional estimator within an iterative-imputation step for one selected feature, while the remaining features are handled classically.}
    \label{fig:quantum_circuit}
    \end{figure}
\subsubsection{State Initialisation: Non-Gaussian Inputs}
\label{subsubsec:state_init}

The first stage of the QNN prepares the quantum register in a non-Gaussian initial state. This choice is essential because the parametrised layers of our model are constructed from fermionic linear optics (FLO) circuits, which are efficiently classically simulable when initialised in Gaussian states \cite{knill2001fermionic, oszmaniec2022fermion}. To access regimes that are challenging for classical simulation, non-Gaussian inputs are therefore required.

We adopt the magic state loader protocol introduced by Oszmaniec \emph{et al.}~\cite{oszmaniec2022fermion}, which prepares entangled four-qubit non-Gaussian states of the form
\begin{equation}
\ket{\phi} = \frac{1}{\sqrt{2}} \left( \ket{0011} + \ket{1100} \right).
\end{equation}


For an $n$-qubit register with $n$ divisible by four, the full initial state is constructed as a tensor product of independent four-qubit blocks,
\begin{equation}
\ket{\psi_{\mathrm{MS}}} = \bigotimes_{k=1}^{n/4} \ket{\phi}_k.
\label{eq:magic_state_loader}
\end{equation}

This preparation ensures that the subsequent FLO-based parametrised circuit operates on a non-Gaussian input, enabling the generation of correlations that are inaccessible to purely Gaussian dynamics. At the same time, the blockwise structure preserves scalability and allows straightforward parallel preparation on hardware.

\subsubsection{Data Loading and Feature Encoding}
\label{subsubsec:data_loading}

Classical data are embedded into the quantum circuit using an angle-encoding scheme based on single-qubit rotations. Specifically, we employ the RY loader introduced in \cite{thakkar2024improved}, which belongs to the broader class of hardware-efficient angle encodings studied in \cite{schuld2018supervised, havlivcek2019supervised, larose2020robust}. The RY encoding layer rotates individual qubit amplitudes but does not map the state back to the Gaussian subspace, so the non-Gaussian character of the initial state is preserved through the data-loading stage. 
Given an $n$-dimensional classical input vector $\mathbf{x} \in \mathbb{R}^n$, the data-loading layer applies a parallel set of $R_Y$ rotations,
\begin{equation}
\ket{\psi_L(\mathbf{x})}
=
\left( \bigotimes_{i=1}^{n} R_y(2\pi x_i) \right) \ket{\psi_{\mathrm{MS}}}.
\end{equation}


This encoding defines a quantum feature map that projects the classical input into a $2^n$-dimensional Hilbert space while maintaining a simple and hardware-efficient gate structure. Crucially, when combined with the subspace-preserving parametrised circuit described below, the data-loading layer provides a mechanism to control the effective dimension of the explored Hilbert space. By adjusting the encoding scheme and circuit depth, one can tune the balance between expressivity and trainability, a key consideration for avoiding barren plateaus and ensuring stable optimisation.

\subsubsection{Parametrised Circuit: Butterfly Architecture}
\label{subsubsec:butterfly}

The trainable core of the QNN is a subspace-preserving parametrised quantum circuit based on the Butterfly architecture proposed in \cite{cherrat2024quantum}, building on the excitation-preserving QNN framework of \cite{landman2022quantum}.
The circuit is composed of layers of two-qubit Reconfigurable Beam Splitter (RBS) gates \cite{johri2021nearest}, defined as
\begin{equation}
\label{eqn:rbs_gate_defn}
\mathsf{RBS}_{i,j}(\theta)
=
e^{-i \frac{\theta}{2} (Y_i \otimes X_j - X_i \otimes Y_j)}
=
\begin{pmatrix}
1 & 0 & 0 & 0 \\
0 & \cos \theta & -\sin \theta & 0 \\
0 & \sin \theta & \cos \theta & 0 \\
0 & 0 & 0 & 1
\end{pmatrix},
\end{equation}
where $\theta$ is a trainable parameter.

RBS gates preserve the total Hamming weight of computational basis states, ensuring that circuit dynamics remain confined to fixed-excitation subspaces of the Hilbert space. The Butterfly architecture arranges these gates in a logarithmic-depth pattern inspired by classical fast-transform networks, such as the Cooley--Tukey fast Fourier transform. For an $n$-qubit system, the circuit has depth $O(\log n)$ and contains $O(n \log n)$ trainable parameters, a substantial reduction compared to densely connected architectures with $O(n^2)$ parameters.

Despite this reduction, successive layers induce global mixing across qubits, allowing the circuit to represent complex correlations. The disjoint gate structure within each layer supports parallel execution and aligns naturally with hardware platforms offering long-range connectivity, such as trapped-ion processors. Finally, the hierarchical structure of the Butterfly circuit directly supports the layer-wise training strategy employed in this work: two independently trained Butterfly circuits of size $n/2$ can be composed into a larger $n$-qubit circuit by adding a final layer of $n/2$ RBS gates that couple the subsystems. This compositional property underpins the training methodology described in the following section.

\subsection{Layer-Wise Training}
\label{subsec:layer_wise}

A central obstacle to scaling quantum neural networks (QNNs) is the prohibitive cost and noise sensitivity associated with training large parametrised circuits entirely on quantum hardware. Circuit depth, coherence time, and measurement budgets impose severe constraints, particularly when gradient-based optimisation is required. To address these challenges, we adopt a \emph{layer-wise training} strategy that combines classical, simulated, and on-hardware optimisation in a structured and principled manner.

The key idea is to exploit the hierarchical structure of the Butterfly architecture and the fact that it only has $O(\log n)$ layers. Rather than optimising all parameters simultaneously, the model is trained progressively: smaller Butterfly subcircuits are trained first, their parameters are fixed, and additional coupling layers are introduced and optimised sequentially. Concretely, we begin by training two independent Butterfly circuits acting on $n/2$ qubits each, either classically or using quantum simulation. Once these subcircuits have converged, their parameters are frozen. A new layer of $n/2$ two-qubit gates is then added to couple the two subcircuits, forming a Butterfly circuit on $n$ qubits. Only the parameters of this newly introduced layer are optimised, while all previously trained layers remain fixed.

This procedure is illustrated schematically in Fig.~\ref{fig:TrainingFlowchart}. By construction, each training stage involves only a small, well-structured subset of parameters, significantly reducing the optimisation complexity and the number of quantum circuit evaluations required. Importantly, because the Butterfly architecture has only $O(\log n)$ layers, the total number of such training stages scales logarithmically with system size.

\tikzstyle{startstop} = [rectangle, rounded corners, 
minimum width = 3cm,
minimum height = 1cm,
text centered,
text width = 0.7\linewidth,
draw=black, 
]

\tikzstyle{arrow} = [thick,->,>=stealth]

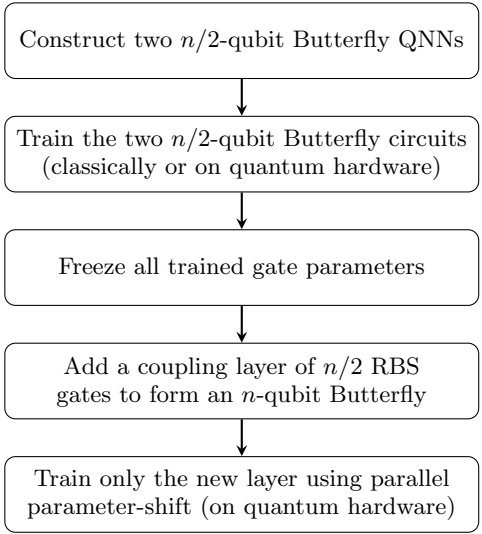
\begin{figure}[!htb]
    \centering
    \begin{tikzpicture}[node distance=1.5cm]

\node (a1) [startstop] {Construct two $n/2$-qubit Butterfly QNNs};
\node (a2) [startstop, below of=a1] {Train the two $n/2$-qubit Butterfly circuits (classically or on quantum hardware)};
\node (a3) [startstop, below of=a2] {Freeze all trained gate parameters};
\node (a4) [startstop, below of=a3] {Add a coupling layer of $n/2$ RBS gates to form an $n$-qubit Butterfly};
\node (a5) [startstop, below of=a4] {Train only the new layer using parallel parameter-shift (on quantum hardware)};


\draw [arrow] (a1) -- (a2);
\draw [arrow] (a2) -- (a3);
\draw [arrow] (a3) -- (a4);
\draw [arrow] (a4) -- (a5);

\end{tikzpicture}
    \caption{
    Layer-wise training workflow for the Butterfly quantum neural network.
    Smaller $n/2$-qubit Butterfly subcircuits are trained first (classically or on quantum hardware) and then frozen.
    A final coupling layer is added to form an $n$-qubit circuit, and only this newly introduced layer is optimised on quantum hardware using parallel parameter-shift, substantially reducing measurement and optimisation cost.
    }
    \label{fig:TrainingFlowchart}
\end{figure}

This layer-wise strategy serves two complementary purposes. First, it drastically limits the number of parameters that must be trained directly on quantum hardware, thereby mitigating the impact of noise and reducing wall-clock training time. Second, it provides a natural interface between classical and quantum resources: early layers can be trained efficiently using classical optimisers or simulators, while later layers---which are most critical for capturing global correlations---are optimised on the quantum processor.

Such progressive or greedy layer-wise optimisation schemes are well established in classical deep learning and block-coordinate optimisation \cite{hinton2006fast,tseng2001convergence}, where they are used to stabilise optimisation, improve convergence, and enable the training of deep or hierarchical models. We note that freezing previously trained layers is a greedy approximation that may not yield globally optimal parameters; an empirical comparison with full fine-tuning is left to future work. In our setting, the same procedure can be applied entirely in classical or simulated regimes, yielding a consistent and well-founded training paradigm across classical, simulated, and hardware-executed quantum neural networks.

\subsection{Parallel Parameter-Shift Rule}
\label{subsec:parallel_param_shift}

Training parametrised quantum circuits relies on estimating gradients of a loss function with respect to circuit parameters. On quantum hardware, this is most commonly achieved using the parameter-shift rule, which yields exact gradients and is compatible with native gate implementations \cite{mitarai2018quantum, schuld2019evaluating}. In generic parametrised quantum circuits, this procedure requires evaluating a distinct shifted circuit for each parameter, leading to a gradient-evaluation cost that scales linearly with the number of trainable parameters. For large quantum neural networks, this measurement overhead quickly becomes prohibitive on NISQ devices \cite{abbas2023quantum}.

Our approach exploits recent advances in the theory of \emph{commuting-block} parametrised quantum circuits, which enable parallel gradient evaluation when circuit structure permits \cite{bowles2023backpropagation, coyle2024training}.

\paragraph{Commuting-block QNNs}

Commuting-block QNNs are parametrised circuits composed of $B$ \emph{blocks} (or layers), such that the overall unitary takes the form
\begin{equation}
\label{eqn:commuting_block_circuits}
\mathcal{U}(\boldsymbol{\theta})
=
\prod_{b=1}^B
\prod_{j=1}^{N_b}
U_b(\boldsymbol{\theta}^b_j)
=
\prod_{b=1}^B
\prod_{j=1}^{N_b}
e^{i\theta^b_j G^b_j}.
\end{equation}
Within each block $b \in \{1,\dots,B\}$, the generators $\{G^b_j\}_{j=1}^{N_b}$ mutually commute, i.e.,
\[
[G^b_i, G^b_j] = 0 \quad \forall i,j.
\]

Assuming without loss of generality that $b=1$, and given a measurement observable $\mathcal{H}$ defining the loss function $\mathcal{L}$, each generator $G_k$ defines a gradient observable
\[
\mathcal{O}_k := [G_k, \mathcal{H}],
\]
such that
\[
i\,\bra{\psi(\boldsymbol{\theta},\boldsymbol{x})}
[G_k,\mathcal{H}]
\ket{\psi(\boldsymbol{\theta},\boldsymbol{x})}
=
\frac{\partial \mathcal{L}}{\partial \theta_k}.
\]
If all generators commute or anticommute with $\mathcal{H}$, i.e.,
\[
[G_k,\mathcal{H}] = 0
\quad \text{or} \quad
\{G_k,\mathcal{H}\} = 0
\quad \forall k,
\]
then the gradient observables $\{\mathcal{O}_k\}$ mutually commute and are therefore simultaneously diagonalisable \cite{bowles2023backpropagation}. As a result, the gradients of all parameters within a block can be extracted in parallel from a single set of circuit evaluations.

\paragraph{Application to the Butterfly architecture}
The Butterfly circuit naturally realises a commuting-block structure. Each of its $\log(n)$ sub-layers consists of Reconfigurable Beam Splitter (RBS) gates acting on disjoint pairs of qubits. Consequently, the generators within each sub-layer mutually commute, and each Butterfly layer constitutes a commuting block in the sense of Eq.~\eqref{eqn:commuting_block_circuits}.

More precisely, the generators of the RBS gate, $G_j \propto Y_{j_1}\otimes X_{j_2} -X_{j_1} \otimes Y_{j_2}$ commute and therefore allow us to use the results of \cite{bowles2023backpropagation} when the measurement Hamiltonian, $\mathcal{H}$ either commutes or anticommutes with $G_j$. If we measure the global version of the observable, $Z^{\otimes n}$, then since $[Z_j^{\otimes 2}, Y_{j_1}\otimes X_{j_2} -X_{j_1} \otimes Y_{j_2}] = 0$, we observe that the measurement operator commutes with the generator. Therefore the condition is satisfied and the parallel parameter-shift rule applies. This is explained in greater detail in \cite{coyle2024training}.

This structural property allows us to apply the parameter-shift rule to all parameters in a Butterfly layer simultaneously. For an individual RBS gate with parameter $\theta_i$, the parameter-shift rule derived in \cite{anselmetti2021local} is given by
\begin{align}
\label{eqn:rbs_parameter_shift}
\frac{\partial f(\boldsymbol{\theta})}{\partial \theta_i}
&=
\left[
f\!\left(\theta_i+\frac{\pi}{4}\right)
-
f\!\left(\theta_i-\frac{\pi}{4}\right)
\right]
\nonumber\\
&\quad
-
\frac{\sqrt{2}-1}{2}
\left[
f\!\left(\theta_i+\frac{\pi}{2}\right)
-
f\!\left(\theta_i-\frac{\pi}{2}\right)
\right].
\end{align}

Exploiting the commuting structure of the Butterfly layer, all parameters $\{\theta_i\}$ in the layer can be shifted simultaneously, yielding
\begin{align}
\label{eqn:rbs_parallel_parameter_shift}
\frac{\partial f(\boldsymbol{\theta})}{\partial \theta_i}
&=
\left[
f\!\left(\boldsymbol{\theta}+\frac{\pi}{4}\right)
-
f\!\left(\boldsymbol{\theta}-\frac{\pi}{4}\right)
\right]
\nonumber\\
&\quad
-
\frac{\sqrt{2}-1}{2}
\left[
f\!\left(\boldsymbol{\theta}+\frac{\pi}{2}\right)
-
f\!\left(\boldsymbol{\theta}-\frac{\pi}{2}\right)
\right],
\qquad \forall i.
\end{align}
Individual gradients are recovered by diagonalising the corresponding commuting gradient observables.

\paragraph{Training complexity}
This parallelised parameter-shift rule drastically reduces the cost of gradient evaluation. For an $n$-qubit Butterfly circuit containing $O(n)$ parameters per layer, the gradients of all parameters can be obtained using a constant number of circuit evaluations, reducing the per-layer gradient cost from $O(n)$ to $O(1)$. Combined with the fact that the Butterfly architecture contains only $O(\log n)$ layers, and with the layer-wise training strategy described in Section~\ref{subsec:layer_wise}, the total number of circuit evaluations per optimisation step scales as $O(\log n)$ (see Table \ref{tab:CircuitScaling}).
This scaling makes gradient-based optimisation of quantum neural networks feasible on current quantum hardware, and is a key enabler of the experimental results presented in the following sections.

\begin{table}[!h]
\centering
\caption{Number of circuit evaluations required to calculate gradients. We compare the number of circuits to be evaluated: first, using the Butterfly circuit with $\frac{1}{2} n \log n$ gates with the naive parameter shift rule (Butterfly+PS); second, the Butterfly circuit with our layer-wise training and parallel parameter shift rule (Butterfly+PPS).}
\label{tab:CircuitScaling}
\begin{tabular}{ccc}
\toprule
\multirow{2}{*}{\;\;Qubits\;\;} & \multicolumn{2}{c}{Circuit Evaluations} \\
                        & \;\;Butterfly + PS  \; \;  & \;\;Butterfly + PPS \; \;  \\
\midrule
8                       & 48                 & 12                 \\
16                      & 128                & 16                 \\
32                      & 320                & 20                 \\
64                      & 768                & 24                 \\
128                     & 1792               & 28                 \\
\bottomrule
\end{tabular}
\end{table}

\paragraph{Summary of the training framework}
Section~\ref{sec:training_qnns} establishes a complete and hardware-compatible framework for training quantum neural networks on near-term quantum processors (Figure \ref{fig:TrainingFlowchart}). By combining non-Gaussian state preparation, subspace-preserving Butterfly architectures, layer-wise optimisation, and parallel gradient evaluation via commuting-block parameter-shift rules, we obtain a QNN training procedure whose cost scales logarithmically with system size. Crucially, each component of this framework is motivated independently by trainability, expressivity, hardware constraints, and their combination yields a coherent optimisation strategy that is feasible on current quantum hardware.

Importantly, the proposed methodology is not tied to a specific learning task or dataset. While the remainder of this work focuses on data imputation as a concrete and challenging application, the same architectural and optimisation principles apply broadly to supervised learning, generative modelling, and hybrid classical-quantum pipelines. In the next section, we instantiate this framework for medical data imputation and empirically evaluate its performance on real-world clinical datasets.

\section{Medical Data Imputation Benchmark}
\label{sec:data_imputation}

\begin{figure*}[!htb]
    \centering
    \includegraphics[scale=0.8]{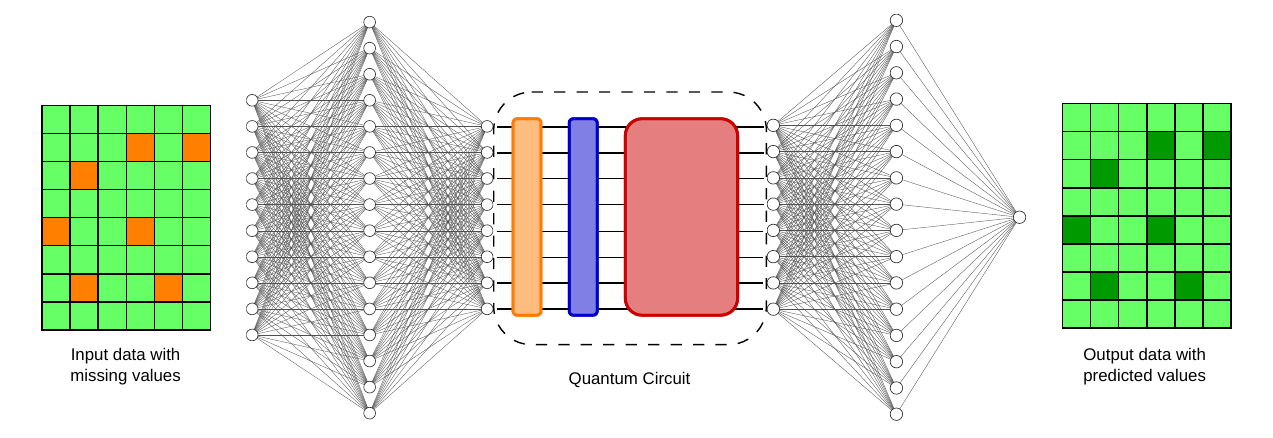}
    \caption{
    Overview of the hybrid classical-quantum imputation pipeline.
    Given a partially observed patient vector, observed features (green) are used as inputs and missing entries (orange) are predicted. The Quantum circuit has multiple components. The first orange box corresponds to the Magic State Loader (Section \ref{subsubsec:state_init}), the second blue box corresponds to the data loader (Section \ref{subsubsec:data_loading}) and the final red box signifies the trainable PQC (Section \ref{subsubsec:butterfly}).
    }
    \label{fig:imputation_pipeline}
\end{figure*}

We now instantiate the training-efficient QNN framework of Section~\ref{sec:training_qnns} on a practically relevant and challenging task: imputing missing values in clinical tabular data.
Clinical datasets routinely suffer from missingness arising from heterogeneous measurement schedules, sensor failures, and incomplete data acquisition, particularly in high-acuity environments such as intensive care units.
These missing entries are not merely a nuisance for data preprocessing; they directly affect the reliability, calibration, and robustness of downstream predictive models used for clinical decision support.

Robust data imputation therefore constitutes a critical component of modern clinical machine learning pipelines.
Effective imputers must capture multivariate dependencies between physiological variables, remain stable under noise and limited data, and generalise reliably across patient populations.
As a result, clinical imputation has become a demanding benchmark for machine learning models, with strong classical baselines based on iterative, ensemble, and deep learning methods.

From the perspective of quantum machine learning, this setting is particularly informative.
Imputation requires learning conditional distributions in moderately high-dimensional feature spaces and is sensitive to optimisation instability and model variance which are precisely the regimes where training efficiency, architectural inductive bias, and robustness to noise are decisive.
By focusing on clinical imputation, we probe not only the expressive capacity of quantum neural networks, but also the practical viability of training them under realistic experimental constraints.

We assess imputation quality indirectly via a clinically meaningful proxy task: binary patient survival prediction.
Following established practice, we evaluate how different imputation strategies affect downstream classifier performance, measured by the area under the ROC curve (AUC).
This protocol enables a fair and application-relevant comparison between classical and hybrid classical-quantum models, and allows us to quantify whether training-efficient QNNs can serve as reliable components within real-world data analysis workflows.

\subsection{Dataset and Prediction Task}
\label{subsec:dataset_task}

Our experiments use the Medical Information Mart for Intensive Care III (MIMIC-III) database \cite{johnson2016mimic}, which contains de-identified electronic health records for intensive care unit (ICU) admissions.
Each patient example is represented by a fixed-length feature vector of physiological measurements (vital signs and derived statistics) following the preprocessing pipeline of \cite{kazdaghli2024improved}.
The downstream task is binary classification of patient outcome (survival), and performance is reported using the AUC metric.

\subsection{Missingness Model}
\label{subsec:missingness_model}

To evaluate imputation methods under controlled conditions, we introduce synthetic missingness using a Missing Completely At Random (MCAR) mechanism \cite{rubin2018multiple}.
Concretely, for each train-test split, we independently mask a fixed fraction of entries uniformly at random across the feature matrix.
Unless stated otherwise, we use a masking rate of $30\%$.
All imputers are fitted using only the observed entries in the training split and are then applied to the corresponding masked test splits. While MCAR is a standard starting point for controlled benchmarking, real clinical datasets typically exhibit MAR or MNAR patterns; robustness under these mechanisms is an important direction for future work.

\subsection{Evaluation Protocol}
\label{subsec:eval_protocol}

Imputation quality is evaluated indirectly via downstream predictive utility, following standard practice in clinical imputation benchmarks \cite{kazdaghli2024improved, shadbahr2023impact}.
For each imputation method, we:
(i) generate a masked dataset according to Section~\ref{subsec:missingness_model},
(ii) impute missing entries using the method under evaluation,
and (iii) train a fixed downstream classifier on the imputed training data and evaluate its AUC on imputed test data.
To capture variability due to random initialisation, stochastic optimisation, and masking randomness, each experiment is repeated over multiple random seeds, and we report the distribution (or mean and standard deviation) of AUC scores.

We now evaluate the proposed hybrid quantum from the perspective of quantum machine learning.

\subsection{Classical Imputation Baselines}
\label{subsec:classical_imputation_baselines}

We compare our hybrid classical-quantum approach against a range of widely used classical imputation methods, spanning simple statistical baselines and stronger iterative/model-based approaches.

\paragraph{Statistical baselines} We include two non-model-based imputers implemented via \texttt{SimpleImputer} in \texttt{scikit-learn}: \textbf{zero imputation}, which replaces missing values with $0$, and \textbf{mean imputation}, which replaces missing values with the empirical mean of the feature computed from observed training entries.

\paragraph{Model-based and iterative imputers} We include \textbf{KNN imputation} (\texttt{KNNImputer},~\texttt{scikit-learn}), which imputes entries using nearest neighbours in feature space; \textbf{MICE (linear)} (\texttt{IterativeImputer},~\texttt{scikit-learn}), using Bayesian ridge regression as the default conditional model~\cite{van2011mice}; \textbf{MissForest}, iterative imputation using random forests~\cite{stekhoven2012missforest}; and \textbf{Deep MICE}, an \texttt{IterativeImputer}-style procedure in which the conditional model is replaced by a neural network to capture non-linear dependencies.
These baselines reflect the spectrum of approaches commonly used in practice and in recent comparative studies \cite{shadbahr2023impact}.
They provide a strong classical reference for assessing the incremental value of the quantum component.

\subsection{Hybrid Classical-Quantum Imputation}
\label{subsec:hybrid_quantum_imputation}

Our approach follows the iterative-imputation philosophy: a conditional model predicts one feature from the others, and the prediction is used to fill the missing entries.
The novelty is that the conditional model is a \emph{hybrid classical-quantum neural network} in which a QNN provides a trainable feature transformation within an otherwise classical network.

To isolate the contribution of the quantum component in a controlled proof-of-concept, we
use a one-feature imputation protocol.
We select one target feature column $j$ to be imputed by the hybrid model, namely the
one with highest predictive importance based on the Gini importance metric.
All remaining feature columns are imputed using a classical method (MissForest in our
experiments).
The hybrid model is then trained to predict feature $j$ from the remaining features,
using only the training split.
At inference time, missing entries of feature $j$ are filled using the hybrid model,
while the other features remain as produced by the classical imputer.
This setting mimics a single conditional-update step of chained-equations imputation \cite{van2011mice}, while keeping the experimental design simple enough to attribute differences in downstream performance to the quantum-enhanced estimator. 
We emphasise that this experimental setup is intended as a controlled diagnostic benchmark to evaluate the training and robustness properties of the proposed quantum–classical models, rather than as a fully optimised production-grade imputation system.

\subsection{Hybrid Model Specification}
\label{subsec:hybrid_model_spec}

We first identify a strong classical reference model by performing an ablation over network width and depth in the fully classical setting.
This study shows that the best-performing classical imputer for the dataset considered has the architecture
$ d \rightarrow 128 \rightarrow 128 \rightarrow 128 \rightarrow 128 \rightarrow 1 $,
which achieves both stable optimisation and the lowest final loss (see Fig.~\ref{fig:loss_trend}).
Notably, increasing the width beyond 128 does not improve performance and in fact leads to degraded convergence, indicating that this architecture represents a suitable capacity for the task.

Starting from this optimal classical baseline, we construct hybrid classical-quantum architectures by replacing the two central hidden layers with a quantum neural network (QNN), yielding models of the form
$d \rightarrow 128 \rightarrow n \rightarrow n \rightarrow 128 \rightarrow 1$ , 
where $n$ denotes the number of qubits in the quantum layer, with $n \in \{8,16,32\}$ in our experiments, while the $128 \rightarrow n $ and $n \rightarrow 128$ remain classical.
The input and output classical layers of width 128 are kept fixed in order to match the representational capacity of the best classical model as closely as possible, while isolating the effect of the quantum component.

By analogy, the classical model with 128 hidden units provides a capacity reference point for the hybrid model, though the function classes are not identical.
Current quantum hardware constraints restrict us to smaller values of $n$, but the performance observed at $n=16$ and $n=32$ provides a controlled diagnostic of how quantum layers function as substitutes for classical hidden layers.
This framing allows us to assess how quantum layers approximate or replace classical hidden representations under realistic qubit constraints and evaluate the effectiveness of the proposed training-efficient QNN framework.

The classical layers use standard non-linear activations (ReLU and tanh), and the QNN layer consists of (i) magic-state initialisation, (ii) an $R_Y$ data loader, and (iii) a Butterfly ansatz with RBS gates.
Training is performed using gradient-based optimisation; when the QNN parameters are trained on hardware, gradients are estimated using the parallel parameter-shift method of Section~\ref{subsec:parallel_param_shift}, optionally within the layer-wise protocol of Section~\ref{subsec:layer_wise}. 
Section~\ref{sec:results} reports the empirical performance of the classical baselines and of the hybrid classical-quantum imputer under simulation and hardware execution.

\section{Results}
\label{sec:results}

We now present the empirical evaluation of the proposed training-efficient quantum neural network (QNN) framework applied to clinical data imputation.
Our results address three core questions central to near-term quantum machine learning:
(i) how hybrid classical-quantum imputers compare to strong classical baselines,
(ii) whether QNNs trained directly on quantum hardware perform comparably to simulation,
and (iii) how performance scales with increasing qubit counts and model expressivity.
All experiments use the MIMIC-III clinical dataset with synthetically introduced missingness and follow the evaluation protocol described in Section~\ref{sec:data_imputation}.
Imputation quality is assessed indirectly through downstream patient survival prediction, measured via the area under the ROC curve (AUC).
Unless otherwise stated, results are reported across multiple random seeds to capture variability arising from model initialisation, imputation stochasticity, and classifier training.

\label{subsec:classical_results}

\subsection{Classical Imputation Baseline}

We begin by establishing rigorous classical reference points.
Table~\ref{tab:ClassicalMethods} summarises the performance of a range of widely-adopted classical imputation methodologies, including simple statistical baselines, distance-based approaches, and iterative model-based techniques.

\begin{table}[!h]
\centering
\caption{Downstream survival prediction AUC for classical imputation baselines and hybrid classical-quantum models on the MIMIC-III dataset with $30\%$ MCAR missingness. The upper bound is set by the fully observed data (no missingness). Bold entries denote hybrid quantum models.}
\label{tab:ClassicalMethods}
\begin{tabular}{lcc}
\toprule
\textbf{Method} & \textbf{Mean AUC} & \textbf{Std} \\
\midrule
Actual (no missingness) & 0.7606 & 0.0016 \\
Deep MICE (NN)          & 0.7176 & 0.0041 \\
\bf{16-Qubit Hybrid DeepImputer} & \bf{0.7147} & \bf{0.0108} \\
MICE (linear)           & 0.7143 & 0.0006 \\
\bf{32-Qubit Hybrid DeepImputer}	& \bf{0.7132} & \bf{0.0161} \\
MissForest              & 0.7127 & 0.0018 \\
KNN Imputation          & 0.7091 & 0.0007 \\
Mean Imputation         & 0.7020 & 0.0007 \\
Zero Imputation         & 0.6998 & 0.0031 \\

\bottomrule
\end{tabular}
\end{table}


The baseline results confirm that simple strategies, such as Zero-filling and Mean Imputation, provide the lowest predictive utility (with $AUC\approx 0.70$), which is expected given their inability to preserve the multivariate correlation inherent in our dataset. 
Among the iterative methods, MissForest and MICE(linear) provide a significant performance increase, yet they remain nearly 5\% lower than the actual data performance ($AUC = 0.7606$).
This difference represents an "information gap" that reflects both the information lost through synthetic missingness and the residual suboptimality of the imputation models themselves.

The most competitive classical benchmark is Deep MICE, which utilises a neural-network-based iterative imputation. 
While it achieves the highest mean-AUC ($AUC=0.7176$), it also exhibits higher variance ($std=0.0041$) compared to the rest of the classical techniques, suggesting that while the neural network architecture captures the more complex non-linear behaviour, it may be more sensitive to the specific missingness distribution.  
These results establish a demanding performance threshold; we therefore use Deep MICE as the primary comparison point in what follows.



\subsection{Hybrid Quantum--Classical Imputation}

We now evaluate the proposed hybrid quantum--classical imputation strategy under realistic hardware and simulation constraints. Our objective in this section is twofold: first, to assess whether quantum-enhanced estimators can match strong classical neural imputers at currently accessible system sizes, and second, to examine how such models behave when extrapolated to larger quantum representations via hardware-compatible inference and simulation.

\paragraph{Matching Classical Baselines}
A primary objective of our evaluation is to determine whether hybrid quantum–classical imputers can achieve performance competitive with strong classical neural baselines at system sizes accessible on current hardware. As established in Table~\ref{tab:ClassicalMethods}, the most demanding classical reference is Deep MICE, which combines iterative imputation with a neural conditional model. We assess competitiveness both in terms of mean predictive performance (AUC) and across-run stability (standard deviation), since robustness under stochastic optimisation and data variability is as clinically relevant as peak accuracy. The results presented in the following paragraphs show that hybrid models trained on 16-qubit systems match Deep MICE in mean AUC while exhibiting consistently lower variance to their fully classical counterparts, suggesting that the structured Butterfly architecture and layer-wise training protocol confer a stabilising inductive bias beyond what is captured by classical network architectures of comparable width.

\paragraph{Feasibility and Qubit Constraints}

A central practical limitation in near-term quantum machine learning is the restricted number of high-fidelity qubits available for training and inference. In our experiments, direct training is performed on IonQ's Forte Enterprise system, which supports up to 36 fully connected trapped-ion qubits. After accounting for calibration overhead and reliability considerations, we restrict on-hardware training to circuits of up to 16 qubits.

To explore larger representations while remaining compatible with hardware constraints, we adopt a hybrid workflow in which training is performed on 16-qubit models, while inference is extended to 32-qubit models using components that are either trained classically or realised by a quantum layer trained via simulation rather than on hardware. This choice reflects a pragmatic trade-off between experimental feasibility and model expressivity. In particular, 16-qubit systems allow stable gradient-based optimisation on current devices, while 32-qubit circuits provide a meaningful intermediate regime between present hardware limits and the idealised regime in which quantum layers would match the width of optimal classical hidden layers.

\paragraph{Quantum Hardware Platform}

All on-device training experiments are conducted on the IonQ Forte Enterprise trapped-ion quantum computer. This platform offers all-to-all qubit connectivity, high-fidelity two-qubit gates, and long coherence times, making it particularly well suited for executing structured circuits such as the Butterfly architecture.
The availability of native long-range interactions enables direct implementation of the RBS gates without compilation overhead, preserving the logarithmic-depth structure of the circuit. Combined with our parallel parameter-shift rule, this allows efficient gradient estimation and stable optimisation despite the presence of hardware noise and finite sampling budgets.




\subsection{Training on Quantum Hardware}
\label{subsec:hardware_training}

\begin{figure}[!htb]
    \centering
    \includegraphics[width=\columnwidth]{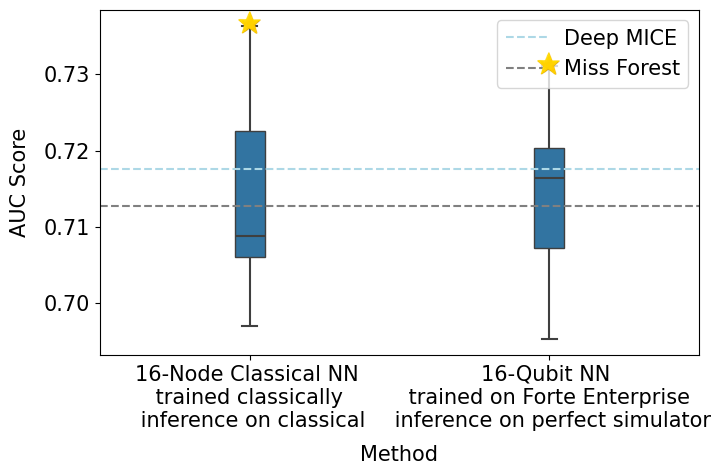}
    \caption{
    Downstream survival prediction AUC for a classical $16 \times 16$ neural network trained classically and a hybrid classical-quantum neural network with a 16-qubit quantum layer trained directly on the IonQ Forte Enterprise trapped-ion processor (16-Qubit Hybrid DeepImputer). Both models share the same overall architecture; inference is performed on a classical simulator in both cases. The hybrid model achieves a higher median AUC with reduced variance across random seeds, demonstrating that on-hardware training via the parallel parameter-shift rule yields models competitive with classical counterparts. 
    }
\label{fig:ClassicalVsQuantum_16qubits_Training}
\end{figure}

The central result of this section is a feasibility demonstration: gradient-based training of a 16-qubit QNN via the parallel parameter-shift rule is viable on current trapped-ion hardware. Prior to this work, the dominant bottleneck preventing on-hardware gradient optimisation at this scale was the $O(n^2)$ circuit evaluation cost of standard parameter-shift, which makes iterative optimisation on real devices prohibitively slow and noise-sensitive beyond modest qubit counts. The framework introduced in Section \ref{sec:training_qnns} reduces this cost to $O(\log n)$, and the experiments reported here confirm that this reduction is sufficient to make direct hardware training tractable at 16 qubits without requiring gradient pruning, zero-order approximations, or simulation fallback.

\begin{figure}[!b]
    \centering
    \includegraphics[width=\columnwidth]{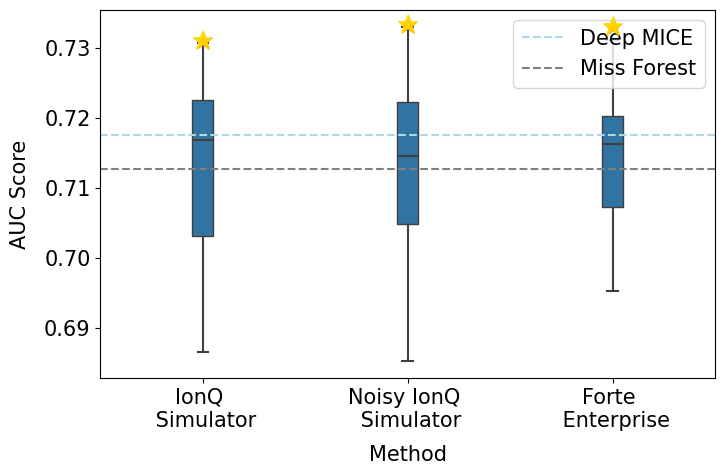}
    \caption{
    Comparison of the 16-Qubit hybrid neural network model performance when the quantum neural network is trained using
    ideal simulation, noisy simulation, and direct execution on IonQ Forte Enterprise hardware. 
    No statistically significant degradation is observed when moving from simulation to hardware. The inference was performed on the classical simulator for all.}
    \label{fig:SimVsHard}
\end{figure}

The QNN is initialised by training two independent 8-qubit subcircuits via classical simulation under the layer-wise protocol of Section \ref{subsec:layer_wise}. The final coupling layer of the full 16-qubit Butterfly circuit is then trained directly on IonQ Forte Enterprise using the parallel parameter-shift rule. 

Figure \ref{fig:ClassicalVsQuantum_16qubits_Training} compares the hardware-trained 16-qubit QNN against a classical 16×16 neural network of equivalent width, serving as an empirical check against performance degradation. Both models operate within a similar AUC range ($\sim$ 0.70 to 0.73), and no measurable AUC penalty is observed for the hardware-trained model. These results confirm that gradient-based training on IonQ Forte Enterprise - facilitated by the parallel parameter-shift rule - is not only feasible but yields models competitive with classical networks of comparable width.
Indeed, the hardware-trained model achieves a marginally higher median AUC ($+1\%$) with lower variance ($-1.6\%$ in standard deviation) relative to the classical reference, suggesting a mild stabilising effect of on-hardware optimisation at this scale.
Figure~\ref{fig:SimVsHard} compares the downstream survival prediction AUC of models trained on an ideal simulator, noisy simulator, and hardware execution on IonQ Forte Enterprise. No statistically significant degradation is observed across these three training conditions, confirming that the $O(\log n)$ training framework is robust to realistic hardware noise at this scale.


\subsection{Scaling to 32 Qubits}
\label{subsec:scaling_32}

\begin{figure}[!thb]
    \centering
    \includegraphics[width=\columnwidth]{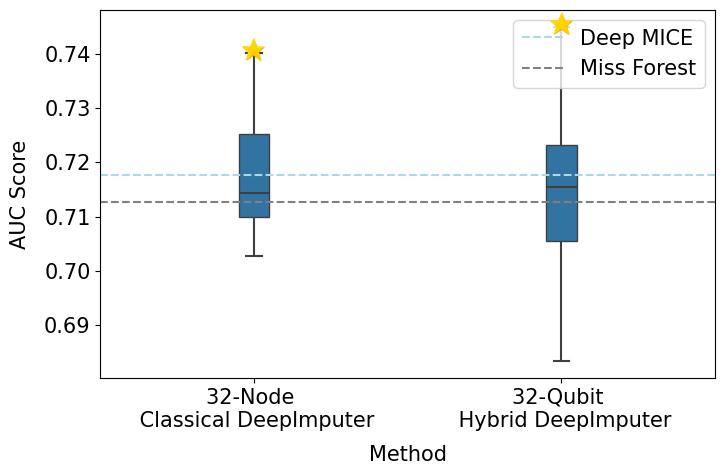}
    \caption{
    Downstream survival prediction AUC for two imputation models at the 32-qubit scale. 32-Node Classical DeepImputer: a fully classical network with central hidden layers of width 32. 32-Qubit Hybrid DeepImputer: the same architecture with central layers replaced by a 32-qubit QNN trained via MPS simulation and evaluated via inference on IonQ Forte Enterprise hardware. Box plots show AUC distributions across multiple random seeds. The hybrid model matches its classical counterpart of equivalent width, confirming that 32-qubit circuit inference on hardware is feasible and yields no performance penalty relative to a classical network of comparable capacity.}
    \label{fig:32qubit}
\end{figure}

\paragraph{Inference with 32-Qubit DeepImputer}
While training at 32 qubits is performed using MPS simulation, inference with the resulting models is executed directly on the IonQ Forte Enterprise hardware, confirming that the trained circuits are hardware-compatible at this scale. Using the one-feature imputation protocol of Section~\ref{subsec:hybrid_quantum_imputation}, we compare two representative models at the 32-qubit scale, \textit{(i) \textbf{32-Node Classical DeepImputer}}: a fully classical neural network with the central hidden layer of dimension $32 \times 32$ 
and \textit{(ii) \textbf{32-Qubit Hybrid DeepImputer}}: a hybrid classical-quantum model 
in which the central $32x32$ layer is realised by a 32-qubit QNN.
For the quantum model, training is performed on 16 qubits using the layer-wise protocol of Section~\ref{subsec:parallel_param_shift}, and the learned subcircuits are composed and extended to 32 qubits. The final 32-qubit layer is trained using an MPS simulator. The classical layers are kept identical across models to ensure comparability.
Figure~\ref{fig:32qubit} summarises the downstream AUC for these two models. The 32-qubit hybrid model achieves performance comparable to its classical counterpart of equivalent width, confirming that a quantum layer trained via MPS simulation can match a classical hidden layer of the same dimension under realistic resource constraints. Crucially, the inference step — executed directly on IonQ Forte Enterprise — introduces no measurable performance penalty, demonstrating that 32-qubit circuits produced by the layer-wise training framework are hardware-compatible at this scale. The remaining classical layers are kept identical across both models to ensure that any performance difference is attributable solely to the quantum component.



\paragraph{Tensor-Network-Based Training and Approximation}

To support training beyond current hardware limits, we employ tensor-network simulations based on matrix product states (MPS) with bond dimension 32. At this setting, simulations can be performed while preserving the dominant entanglement structure induced by the magic-state initialisation and Butterfly ansatz, providing a practical and reproducible training environment for circuits larger than those directly accessible on current hardware.

\subsection{Scaling Outlook: Representational Capacity}
\label{subsec:scaling_outlook}

\begin{figure}[!htb]
    \centering
    \includegraphics[width=\columnwidth]{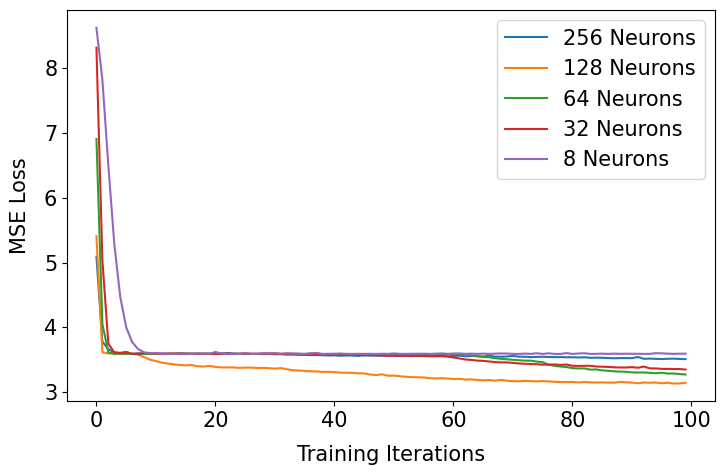}
\caption{Training convergence of classical neural networks with varying hidden-layer widths. Mean Squared Error (MSE) as a function of training iterations for networks with 8, 32, 64, 128, and 256 hidden units. Performance improves monotonically up to 128 units, while further expansion to 256 units leads to degraded convergence and a higher final loss. This saturation behaviour motivates the choice of 128 units as the optimal classical baseline capacity and, by the architectural correspondence between classical and hybrid models, identifies 128 qubits as the target scale for fully quantum-enhanced imputation.}
    \label{fig:loss_trend}
\end{figure}

To conclude our empirical analysis, we evaluate the representational capacity required to accurately model the underlying problem of clinical data distributions. As shown in Figure~\ref{fig:loss_trend}, we performed an ablation study on the hidden-layer width of classical neural networks to identify the point of performance saturation. The training convergence curves demonstrate that while model performance improves monotonically as the width increases from 8 to 128 neurons, further expansion to 256 neurons leads to degraded convergence and a higher final Mean Squared Error (MSE) loss. This empirical evidence suggests that a classical hidden layer of 128 units represents the optimal capacity for this specific imputation task, beyond which the model appears to overfit or encounter optimisation instabilities.

Based on the architectural correspondence between classical and hybrid models (established in Section~\ref{subsec:hybrid_model_spec}), where the classical baseline serves as a limiting case for the hybrid model, we conjecture that a QNN would require approximately 128 qubits to fully match the expressive power of this optimal classical hidden representation. While full-dataset imputation with 128-qubit layers remains a future milestone, the favourable scaling trends observed in 32-qubit inference and the performance stability of our hardware-trained models provide a concrete pathway toward surpassing state-of-the-art classical neural imputers as hardware platforms mature.

\section{Conclusions}


This work introduces a hardware-compatible training framework for gradient-based optimisation of quantum neural networks on near-term processors. By combining a structured, logarithmic-depth Butterfly architecture with layer-wise training and a parallelised parameter-shift rule that exploits the commuting structure within each layer, the number of distinct circuit evaluations per optimisation step scales as $O(\log n)$ in the number of qubits — down from $O(n^2)$ for generic architectures — and remains constant regardless of layer width. This reduction constitutes the primary contribution of the framework and is the enabling condition for the hardware training results reported here.

Our empirical evaluation on the MIMIC-III clinical data imputation task confirms that hardware-trained 16-qubit QNNs can be successfully optimised via gradient-based methods directly on the IonQ Forte Enterprise trapped-ion system, with no performance degradation relative to simulation or classical networks of comparable width, and with improved optimisation stability and reduced variance across runs. 32-qubit inference on IonQ Forte Enterprise confirms hardware compatibility at this scale, with the hybrid model matching its classical counterpart of equivalent width. 
MPS simulations with bond dimension 32 yield performance comparable to fully trained classical counterparts, confirming a sufficiently faithful training approximation. Our ablation study identifies 128 qubits as the target capacity for this clinical task, matching the performance saturation point of optimal classical baselines.

\newcommand{\comment}[1]{}
\comment{
The present results nonetheless operate within a controlled one-feature imputation protocol, and the hybrid models rely on classically imputed features as inputs, which constrains the extent to which quantum correlations propagate across the full feature space. Fully quantum-enhanced chained-equations imputation, in which all conditional models are implemented using QNNs, would require hundreds of high-fidelity qubits to match the capacity of optimal classical networks — a regime beyond the reach of current hardware. 
Nevertheless, the favourable scaling trends observed in tensor-network simulations and 32-qubit inference experiments provide evidence that these limitations are primarily technological rather than fundamental.}

As hardware platforms mature, the same training-efficient framework introduced here can be extended to full-dataset imputation, providing a pathway toward real-world clinical data analysis workflows.

\clearpage
\bibliography{apssamp}

@PREAMBLE{
 "\providecommand{\noopsort}[1]{}" 
 # "\providecommand{\singleletter}[1]{#1}%" 
}

@article{kazdaghli2024improved,
  title={Improved clinical data imputation via classical and quantum determinantal point processes},
  author={Kazdaghli, Skander and Kerenidis, Iordanis and Kieckbusch, Jens and Teare, Philip},
  journal={Elife},
  volume={12},
  pages={RP89947},
  year={2024},
  publisher={eLife Sciences Publications Limited}
}

@article{cherrat2024quantum,
  title={Quantum vision transformers},
  author={Cherrat, El Amine and Kerenidis, Iordanis and Mathur, Natansh and Landman, Jonas and Strahm, Martin and Li, Yun Yvonna},
  journal={Quantum},
  volume={8},
  number={arXiv: 2209.08167},
  pages={1265},
  year={2024}
}

@article{landman2022quantum,
  title={Quantum methods for neural networks and application to medical image classification},
  author={Landman, Jonas and Mathur, Natansh and Li, Yun Yvonna and Strahm, Martin and Kazdaghli, Skander and Prakash, Anupam and Kerenidis, Iordanis},
  journal={Quantum},
  volume={6},
  pages={881},
  year={2022},
  publisher={Verein zur F{\"o}rderung des Open Access Publizierens in den Quantenwissenschaften}
}

@article{coyle2024training,
  title={Training-efficient density quantum machine learning},
  author={Coyle, Brian and Raj, Snehal and Mathur, Natansh and Cherrat, El Amine and Jain, Nishant and Kazdaghli, Skander and Kerenidis, Iordanis},
  journal={arXiv preprint arXiv:2405.20237},
  year={2024}
}

@article{bowles2023backpropagation,
  title={Backpropagation scaling in parameterised quantum circuits},
  author={Bowles, Joseph and Wierichs, David and Park, Chae-Yeun},
  journal={arXiv preprint arXiv:2306.14962},
  year={2023}
}

@article{oszmaniec2022fermion,
  title={Fermion sampling: a robust quantum computational advantage scheme using fermionic linear optics and magic input states},
  author={Oszmaniec, Micha{\l} and Dangniam, Ninnat and Morales, Mauro ES and Zimbor{\'a}s, Zolt{\'a}n},
  journal={PRX Quantum},
  volume={3},
  number={2},
  pages={020328},
  year={2022},
  publisher={APS}
}

@article{mitarai2018quantum,
  title={Quantum circuit learning},
  author={Mitarai, Kosuke and Negoro, Makoto and Kitagawa, Masahiro and Fujii, Keisuke},
  journal={Physical Review A},
  volume={98},
  number={3},
  pages={032309},
  year={2018},
  publisher={APS}
}

@article{schuld2019evaluating,
  title={Evaluating analytic gradients on quantum hardware},
  author={Schuld, Maria and Bergholm, Ville and Gogolin, Christian and Izaac, Josh and Killoran, Nathan},
  journal={Physical Review A},
  volume={99},
  number={3},
  pages={032331},
  year={2019},
  publisher={APS}
}

@article{thakkar2024improved,
  title={Improved financial forecasting via quantum machine learning},
  author={Thakkar, Sohum and Kazdaghli, Skander and Mathur, Natansh and Kerenidis, Iordanis and Ferreira--Martins, Andr{\'e} J and Brito, Samurai},
  journal={Quantum Machine Intelligence},
  volume={6},
  number={1},
  pages={27},
  year={2024},
  publisher={Springer}
}

@article{johnson2016mimic,
  title={MIMIC-III, a freely accessible critical care database},
  author={Johnson, Alistair EW and Pollard, Tom J and Shen, Lu and Lehman, Li-wei H and Feng, Mengling and Ghassemi, Mohammad and Moody, Benjamin and Szolovits, Peter and Anthony Celi, Leo and Mark, Roger G},
  journal={Scientific data},
  volume={3},
  number={1},
  pages={1--9},
  year={2016},
  publisher={Nature Publishing Group}
}

@article{van2011mice,
  title={mice: Multivariate imputation by chained equations in R},
  author={Van Buuren, Stef and Groothuis-Oudshoorn, Karin},
  journal={Journal of statistical software},
  volume={45},
  pages={1--67},
  year={2011}
}

@article{stekhoven2012missforest,
  title={MissForest—non-parametric missing value imputation for mixed-type data},
  author={Stekhoven, Daniel J and B{\"u}hlmann, Peter},
  journal={Bioinformatics},
  volume={28},
  number={1},
  pages={112--118},
  year={2012},
  publisher={Oxford University Press}
}

@article{shadbahr2023impact,
  title={The impact of imputation quality on machine learning classifiers for datasets with missing values},
  author={Shadbahr, Tolou and Roberts, Michael and Stanczuk, Jan and Gilbey, Julian and Teare, Philip and Dittmer, S{\"o}ren and Thorpe, Matthew and Torn{\'e}, Ramon Vi{\~n}as and Sala, Evis and Li{\'o}, Pietro and others},
  journal={Communications medicine},
  volume={3},
  number={1},
  pages={139},
  year={2023},
  publisher={Nature Publishing Group UK London}
}

@article{johri2021nearest,
  title={Nearest centroid classification on a trapped ion quantum computer},
  author={Johri, Sonika and Debnath, Shantanu and Mocherla, Avinash and Singk, Alexandros and Prakash, Anupam and Kim, Jungsang and Kerenidis, Iordanis},
  journal={npj Quantum Information},
  volume={7},
  number={1},
  pages={122},
  year={2021},
  publisher={Nature Publishing Group UK London}
}

@article{kerenidis2022quantum,
  title={Quantum machine learning with subspace states},
  author={Kerenidis, Iordanis and Prakash, Anupam},
  journal={arXiv preprint arXiv:2202.00054},
  year={2022}
}

@article{anselmetti2021local,
  title     = {Local, expressive, quantum-number-preserving VQE ans{\"a}tze for fermionic systems},
  author    = {Anselmetti, Gian-Luca R and Wierichs, David and Gogolin, Christian and Parrish, Robert M},
  journal   = {New Journal of Physics},
  volume    = {23},
  number    = {11},
  pages     = {113010},
  year      = {2021},
  publisher = {IOP Publishing}
}

@article{schuld2018supervised,
  title={Supervised learning with quantum computers},
  author={Schuld, Maria and Petruccione, Francesco},
  journal={Quantum science and technology},
  volume={17},
  year={2018},
  publisher={Springer}
}

@article{larose2020robust,
  title={Robust data encodings for quantum classifiers},
  author={LaRose, Ryan and Coyle, Brian},
  journal={arXiv preprint arXiv:2003.01695},
  year={2020}
}

@article{havlivcek2019supervised,
  title={Supervised learning with quantum-enhanced feature spaces},
  author={Havl{\'\i}{\v{c}}ek, Vojt{\v{e}}ch and C{\'o}rcoles, Antonio D and Temme, Kristan and Harrow, Aram W and Kandala, Abhinav and Chow, Jerry M and Gambetta, Jay M},
  journal={Nature},
  volume={567},
  number={7747},
  pages={209--212},
  year={2019},
  publisher={Nature Publishing Group UK London}
}

@article{knill2001fermionic,
  title={Fermionic linear optics and matchgates},
  author={Knill, Emanuel},
  journal={arXiv preprint quant-ph/0108033},
  year={2001}
}

@article{sterne2009multiple,
  title={Multiple imputation for missing data in epidemiological and clinical research: potential and pitfalls},
  author={Sterne, Jonathan AC and White, Ian R and Carlin, John B and Spratt, Michael and Royston, Patrick and Kenward, Michael G and Wood, Angela M and Carpenter, James R},
  journal={Bmj},
  volume={338},
  year={2009},
  publisher={British Medical Journal Publishing Group}
}

@incollection{rubin2018multiple,
  title={Multiple imputation},
  author={Rubin, Donald B},
  booktitle={Flexible imputation of missing data, second edition},
  pages={29--62},
  year={2018},
  publisher={Chapman and Hall/CRC}
}

@article{abbas2023quantum,
  title={On quantum backpropagation, information reuse, and cheating measurement collapse},
  author={Abbas, Amira and King, Robbie and Huang, Hsin-Yuan and Huggins, William J and Movassagh, Ramis and Gilboa, Dar and McClean, Jarrod},
  journal={Advances in Neural Information Processing Systems},
  volume={36},
  pages={44792--44819},
  year={2023}
}

@article{mcclean2018barren,
  title={Barren plateaus in quantum neural network training landscapes},
  author={McClean, Jarrod R and Boixo, Sergio and Smelyanskiy, Vadim N and Babbush, Ryan and Neven, Hartmut},
  journal={Nature Communications},
  volume={9},
  number={1},
  pages={4812},
  year={2018},
  publisher={Nature Publishing Group}
}

@article{cerezo2021cost,
  title={Cost function dependent barren plateaus in shallow parametrized quantum circuits},
  author={Cerezo, M and Sone, A and Volkoff, T and Cincio, L and Coles, PJ},
  journal={Nature Communications},
  volume={12},
  number={1},
  pages={1791},
  year={2021},
  publisher={Nature Publishing Group}
}

@article{monbroussou2025trainability,
  title={Trainability and expressivity of hamming-weight preserving quantum circuits for machine learning},
  author={Monbroussou, L{\'e}o and Mamon, Eliott Z and Landman, Jonas and Grilo, Alex B and Kukla, Romain and Kashefi, Elham},
  journal={Quantum},
  volume={9},
  pages={1745},
  year={2025},
  publisher={Verein zur F{\"o}rderung des Open Access Publizierens in den Quantenwissenschaften}
}

@article{hinton2006fast,
  title={A fast learning algorithm for deep belief nets},
  author={Hinton, Geoffrey E and Osindero, Simon and Teh, Yee-Whye},
  journal={Neural computation},
  volume={18},
  number={7},
  pages={1527--1554},
  year={2006},
  publisher={MIT Press}
}

@article{tseng2001convergence,
  title={Convergence of a block coordinate descent method for nondifferentiable minimization},
  author={Tseng, Paul},
  journal={Journal of optimization theory and applications},
  volume={109},
  number={3},
  pages={475--494},
  year={2001},
  publisher={Springer}
}

@inproceedings{Wang2022,
  title     = {{QOC}: Quantum On-Chip Training with Parameter Shift and Gradient Pruning},
  author    = {Wang, Hanrui and Li, Zirui and Gu, Jiaqi and Ding, Yongshan and Pan, David Z. and Han, Song},
  booktitle = {Proceedings of the 59th ACM/IEEE Design Automation Conference (DAC)},
  pages     = {655--660},
  year      = {2022},
  doi       = {10.1145/3489517.3530495},
  note      = {arXiv:2202.13239}
}

@article{kverne2026wsbd,
  title={WSBD: Freezing-Based Optimizer for Quantum Neural Networks},
  author={Kverne, Christopher and Akewar, Mayur and Huo, Yuqian and Patel, Tirthak and Bhimani, Janki},
  journal={arXiv preprint arXiv:2602.11383},
  year={2026}
}

\end{document}